\documentclass[a4paper]{jpconf}
\usepackage[english]{babel}
\usepackage[T1]{fontenc}

\usepackage{graphicx}
\graphicspath{{./}{Graphics/}}

\usepackage{booktabs}
\usepackage{array} 
\newcolumntype{L}[1]{>{\raggedright\let\newline\\\arraybackslash\hspace{0pt}}m{#1}}
\newcolumntype{C}[1]{>{\centering\let\newline\\\arraybackslash\hspace{0pt}}m{#1}}
\newcolumntype{R}[1]{>{\raggedleft\let\newline\\\arraybackslash\hspace{0pt}}m{#1}}

\usepackage{custom}
\usepackage{url}
\usepackage{hyperref}
\begin{document}

\newcommand{\blue}[1]{\textcolor{blue}{#1}}
\newcommand{\red}[1]{\textcolor{red}{#1}}

\title{Emittance-preserving acceleration of high-quality positron beams using warm plasma filaments}

\author{S Diederichs$^{1}$, C Benedetti$^{2}$, E Esarey$^{2}$, A Sinn$^{1}$, J Osterhoff$^{1,3}$, C B Schroeder$^{2,4}$, and M Thévenet$^{1}$}

\address{$^{1}$Deutsches Elektronen-Synchrotron DESY, Notkestr. 85, 22607 Hamburg, Germany}
\address{$^{2}$Lawrence Berkeley National Laboratory, Berkeley, California 94720, USA}
\address{$^3$Universität Hamburg, Mittelweg 177, 20148 Hamburg, Germany}
\address{$^{4}$Department of Nuclear Engineering, University of California, Berkeley, California 94720, USA}

\begin{abstract}
Preserving the quality of positron beams in plasma-based accelerators, where wakefields are generated in electron filaments, is challenging. These wakefields are characterized by transversely non-linear focusing fields and non-uniform accelerating fields.
However, a non-zero plasma temperature linearizes the transverse wakefield within the central region of the electron filament. In this study, we employ 3D particle-in-cell simulations with mesh refinement to demonstrate that beams with emittances on the order of tens of nanometers are contained within the linearized region of the transverse wakefield. This enables emittance preservation to one percent, while positron beams with the same charge and micrometer emittances, which sample the non-linear part of the transverse wakefield, experience a relative emittance growth of ten percent. Additionally, we observe a significant reduction in the growth rate of the slice energy spread for the tens of nanometers emittance beams in comparison to the micrometer emittance beams.
The utilization of warm plasmas in conjunction with low-emittance beams opens up new avenues for enhancing the beam quality across various plasma-based positron acceleration approaches.
\end{abstract}

\section{Introduction}

Plasma-based accelerators can produce large accelerating fields and are considered a promising technology for a future linear electron-positron collider.
Achieving high beam quality is paramount for the success of a linear collider, as it directly impacts the attainable luminosity. Plasma accelerators have demonstrated efficient, high-quality electron acceleration, primarily utilizing the so-called blowout regime~\cite{Litos:2014, Lindstrom:2021}. The acceleration of high-quality positron bunches presents a more difficult challenge due to the inherent asymmetric response of the plasma~\cite{Lee:2001, Hogan:2003, Muggli:2008, Corde:2015}.

To create plasma fields that simultaneously accelerate and focus positrons, a region of high electron density is essential. Typically, such a high-electron-density region exhibits transversely non-linear focusing and non-uniform accelerating wakefields that vary with the longitudinal position, potentially compromising the positron beam quality.
Several schemes have been proposed for generating the high-density electron filaments required for positron acceleration. These approaches include plasma columns~\cite{Diederichs:2019, Diederichs:2020, Diederichs:2022:DBS, Diederichs:2022:WBS}, using the posterior region of a blowout wake~\cite{Lotov:2007, Zhou:2022, Liu:2022, Wang:2021}, quasi-hollow plasma channels~\cite{Silva:2021}, or hollow core plasma channels operating in the nonlinear regime~\cite{Zhou:2021}.

In most of these schemes, plasmas are created through optical field ionization, as is the case for plasma columns~\cite{Green:2014, Shalloo:2018, Picksley:2020}, or hollow-core plasma channels~\cite{Gessner:NatCom:2016}. Alternatively, homogeneous plasmas can be generated through electrical discharges. In both methods, the plasma electrons typically exhibit temperatures ranging from one to tens of electronvolts (eV)~\cite{Ehrlich:1996, Bobrova:2001, Shalloo:2018}.

Recently, a non-zero plasma temperature was found to have significant effects on positron acceleration in electron filaments~\cite{Diederichs:2023}: it broadens the electron filament, linearizes a small fraction of the transverse wakefield within the filament, and decreases the variance of focusing field along the bunch longitudinally. As a consequence, temperature reduces the emittance growth and slice energy spread of positrons accelerated in an electron filament.
As the beam was larger than the linear region of the transverse wakefields a significant fraction of the positrons sampled nonlinear transverse wakefields, causing the beam emittance to grow significantly~\cite{Diederichs:2023}.

In this work, we show that for a collider-relevant normalized emittance of tens of nanometers, the matched spot size of the positron bunch is small enough to only sample the linear region of the transverse wakefield in a warm electron filament, allowing for emittance preservation to the percent level, while maintaining the same charge. Simultaneously, a beam with collider-relevant emittance samples a smaller region of the transversely non-uniform accelerating field, better maintaining its slice energy spread.

\section{Results and Discussion}

Plasma columns offer a promising avenue for achieving high-quality and stable positron acceleration~\cite{Diederichs:2019, Diederichs:2020, Diederichs:2022:DBS, Diederichs:2022:WBS}. In this scheme, an electron drive beam propagates along the central axis of a finite-radius plasma column. When the electron beam generates a wake in the blowout regime characterized by a blowout radius larger than the plasma column radius, the trajectories of plasma electrons in the sheath are perturbed due to the absence of ions beyond the column's boundary. Consequently, the focusing force acting on these plasma sheath electrons decreases, causing them to return towards the propagation axis in the form of an elongated filament located at a specific distance behind the driver. This electron filament simultaneously provides accelerating and focusing fields for positron beams.

The scheme is demonstrated through simulations using HiPACE\texttt{++}\,\cite{Diederichs:2022:CPC}, employing Gaussian drive and witness beams, and a plasma column with a radius of $R_p = 2.5\,k_p^{-1}$. Here, $k_p^{-1} = c / \omega_p$ is the plasma skin depth and $\omega_p = \sqrt{4\pi n_0 e^2 / m_e}$ (in CGS units) is the plasma frequency with $n_0$ the background plasma density, $e$ the elementary charge, $c$ the speed of light, and $m_e$ the electron mass. The drive beam has a peak density of $n_{b,d} = 200\,n_0$. It is bi-Gaussian with rms sizes $\sigma_{x/y, d} = 0.1\,k_p^{-1}$ and $\sigma_{z,d} = 1.41\,k_p^{-1}$ and is located at the origin. It has an energy of $5\,$GeV ($\gamma_d = 10000$) per electron and its emittance is matched to the blowout wake. The witness beam has a peak density of $n_{b,w} = 500\,n_0$, with rms sizes $\sigma_{x/y, w} = 0.025\,k_p^{-1}$ and $\sigma_{z,w} = 0.5\,k_p^{-1}$ and is located at $\zeta_w\approx-11.6\,k_p^{-1}$ behind the driver, where $\zeta = z - ct$ is the co-moving variable. It has an energy of $1\,$GeV ($\gamma_w = 2000$) per positron, a normalized emittance of $\epsilon_w = 0.1\,k_p^{-1}$, and is matched to the nonlinear focusing wake~\cite{Diederichs:2019}.
The drive and witness beams are modelled by $10^8$ and $2\times10^8$ macro-particles, respectively. The plasma electrons and ions are modeled with 900 and 1 macro-particles per cell, respectively. The large discrepancy of macro-particles of the plasma electrons and ions is owing to the fact that the electron sheath must be precisely resolved. The plasma electrons have a temperature of $50\,$eV and the ions are assumed to be cold.

In these simulations, the computational domain spans $(-16, 16) \times (-16,16) \times (-14, 6)\,k_p^{-3}$ in $x \times y \times \zeta$. The mesh resolution is $0.0078 \times 0.0078 \times 0.0013\,k_p^{-3}$. To resolve small emittance beams, we employ a refined grid that spans $(-0.125, 0.125) \times (-0.125,0.125) \times (-14, 6)\,k_p^{-3}$ in $x \times y \times \zeta$. The mesh resolution of the refined grid is $(2.4 \times 10^{-4}) \times (2.4 \times 10^{-4}) \times 0.0013\,k_p^{-3}$. The full numerical settings can be found in the input scripts, see Sec.~\ref{sec:data_availability}.

\begin{figure}
  \centering
  \includegraphics[width=\columnwidth]{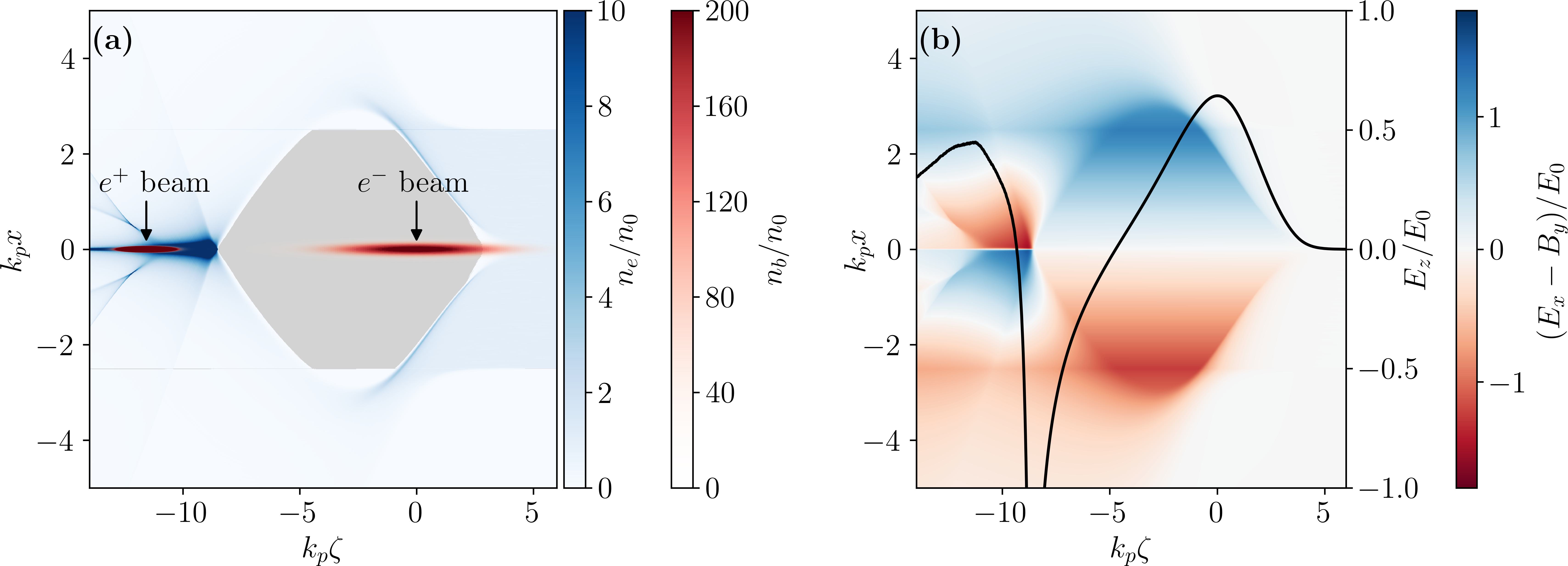}
  \caption{Schematic of positron acceleration in a plasma column. 
	(a) Drive and witness beam densities (red), normalized plasma density (blue), and ion background density (grey) in the $x$-$\zeta$-plane. The drive beam excites a plasma wake with a blowout radius larger than the column radius, thereby generating an intense on-axis electron filament at $-14 < k_p \zeta < -9$. (b) Transverse wakefield in the $x$-$\zeta$-plane (red-blue) and on-axis longitudinal wakefield (black line) along the co-moving variable $\zeta$. The electron filament generates an accelerating and focusing field structure for positrons. }\label{fig:scheme}
\end{figure}

The scheme is illustrated in Figure~\ref{fig:scheme}. The normalized electron density in the $x$-$\zeta$-plane is shown in (a), where the blue, grey and red color scale denote the electron density, the ion background density, and the electron drive and positron witness beam densities, respectively. In Fig.~\ref{fig:scheme}~(b), the resulting transverse wakefield and a lineout of the on-axis longitudinal wakefield (indicated by the black line) are displayed. The on-axis electron filament, spanning from $-14 < k_p \zeta < -10$, generates both accelerating and focusing fields essential for positron beams.

\begin{figure}
  \centering
  \includegraphics[width=\columnwidth]{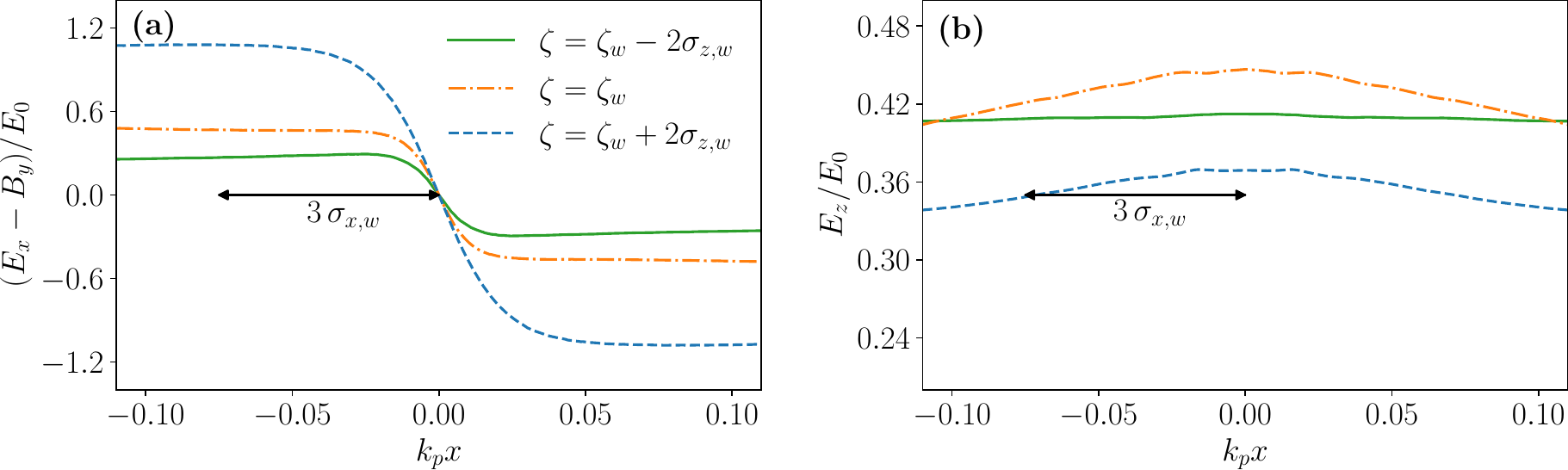}
  \caption{Transverse (a) and longitudinal (b) wakefields along the transverse coordinate $k_px$ at the tail (blue lines), center (orange lines), and head (green lines) of the witness bunch for a plasma with $50$\,eV (solid lines). The transverse extent of the witness bunch (black arrows) at $\epsilon_x = 0.1\,k_p^{-1}$ and $\gamma_w=2000$ exceeds the linear region of the transverse wakefields.}\label{fig:loaded_fields}
\end{figure}

In the case of a cold plasma, the transverse wakefield within the electron filament closely resembles a step-like profile along the transverse coordinate, as previously discussed~\cite{Diederichs:2019}.
Recently, research has revealed that the presence of an electron temperature in the range of tens of eV has a smoothing effect on the step-like transverse wakefield. This results in the emergence of a linear region centered around the zero-crossing point, and reducing slice-to-slice variations in the focusing gradient along the bunch~\cite{Diederichs:2023}. Furthermore, the electron temperature has the effect of transversely flattening the accelerating field.
The impact of these temperature-induced effects on the transverse wakefield $E_x-B_y$ and the longitudinal wakefield $E_z$ along the transverse coordinate $x$ is depicted in Fig.~\ref{fig:loaded_fields}~(a) and (b), respectively. In the figure, the solid green line, the dash-dotted orange line, and the dashed blue line represent the fields at three positions within the positron bunch: the tail (defined as $\zeta_w - 2\sigma_{z,w}$), the center ($\zeta_w$), and the head ($\zeta_w + 2\sigma_{z,w}$), respectively.
As shown in the figure, the transverse size of the bunch (indicated by the black arrow) exceeds both the linear region of the focusing field and the nearly uniform region of the accelerating field. The sampling of the non-linear focusing and non-uniform accelerating field results in emittance growth and induces a slice energy spread during the acceleration process.

Given the wakefield structure described, one could expect that a lower emittance beam with a matched spot size small enough to fit into the linear region of the transverse wakefield and the close-to-uniform region of the longitudinal wakefield could preserve both the emittance and the slice energy spread. This expectation holds under the premise that the wakefield structure remains intact when the positron beam density exceeds that of the electron filament.
Owing to the utilization of mesh refinement, we can, for the first time, model this low-emittance beam scenario to an excellent level of numerical convergence. Reducing the positron beam normalized emittance to $\epsilon_{x/y} = 0.002\,k_p^{-1}$, which corresponds to a matched spot size of $\sigma_{x/y, w} = 0.0026\,k_p^{-1}$ and a peak density of $n_{b,w} \approx 46000\,n_0$, we find that the wakefield structure is almost unchanged. Thereby, the matched spot size fits into the linear region of the focusing wakefield, enabling emittance preservation.

\begin{figure}
  \centering
  \includegraphics[width=\columnwidth]{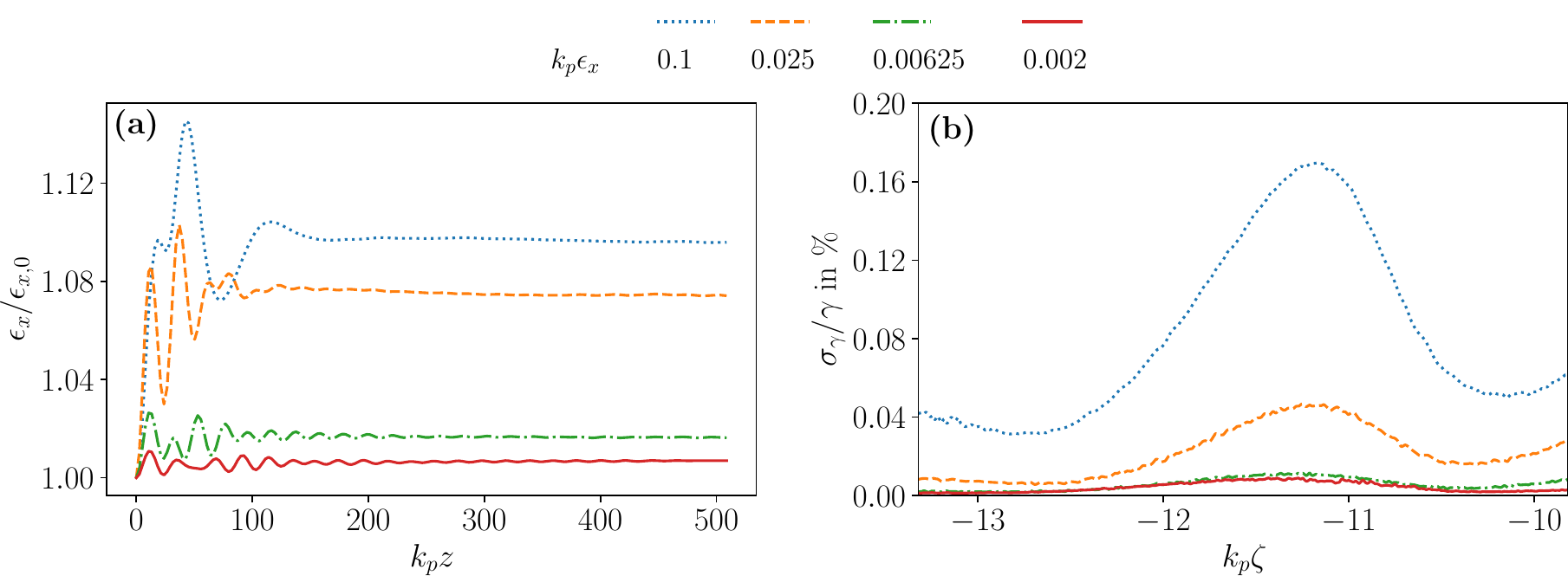}
  \caption{(a) Relative growth of the projected emittance and (b) slice energy spread for initial beam normalized emittances between $k_p \epsilon_x =0.1$ (dotted blue line) and $0.002$ (solid red line). A smaller emittance is better preserved and yields a lower slice energy spread. Note that $k_p \epsilon_x = 0.002 $ corresponds to 33~nm at $n_0=10^{17}$~cm$^{-3}$.}
  \label{fig:emittance_energy_spread}
\end{figure}

The evolution of the relative emittance growth and the slice energy spread for various initial emittance values is depicted in Fig.~\ref{fig:emittance_energy_spread}~(a) and (b), respectively. In the initial setup featuring a positron beam with $\epsilon_{x/y} = 0.1\,k_p^{-1}$ (dotted blue line), the relative emittance growth reaches approximately $10\%$. As expected, the lowest emittance beam (solid red line) exhibits both the lowest relative emittance growth and the smallest slice energy spread. At a background plasma density of $n_0 = 10^{17}\,$cm$^{-3}$ the emittance of $\epsilon_{x/y} = 0.002\,k_p^{-1}$ corresponds to $\approx 33\,$nm and is preserved to the percent-level. Consequently, a lower emittance beam, a requirement for applications in linear colliders, offers superior emittance preservation and reduced slice energy spread while maintaining the same charge. This, in turn, significantly enhances the achievable luminosity for positron acceleration in a plasma column in comparison to previous results~\cite{Diederichs:2019, Diederichs:2020}. Recent work has introduced luminosity per power as a metric to compare plasma-based positron acceleration schemes~\cite{Cao:2023}. In this work, we have used 25 times smaller emittances than in \cite{Diederichs:2019,Diederichs:2020} (while maintaining the other parameters), suggesting that a 25 times higher luminosity per power than reported in \cite{Cao:2023} is achievable. 

\section{Conclusion}
In this study, we have employed quasi-static PIC simulations with mesh refinement to demonstrate the preservation of positron beam emittance at the tens of nanometer scale, achieving a relative emittance growth below one percent while accelerating the beam within a warm plasma column. The small size of the matched spot for such a tens-of-nanometer emittance beam enables efficient utilization of the linear region within the transverse wakefields generated in the warm plasma column, facilitating emittance preservation. Furthermore, this configuration minimizes the fraction of the beam interacting with the non-uniform accelerating field, resulting in a reduced slice energy spread. 
These promising findings mark a significant advancement in positron acceleration within a plasma column. Notably, the luminosity per power in this study surpasses the previously reported value in \cite{Cao:2023} by a factor of 25.
We anticipate that these results will have broader applicability to other positron acceleration techniques employing electron filaments~\cite{Zhou:2022, Wang:2021, Zhou:2021}. In summary, the linearization of the focusing field due to temperature effects offers a pathway for achieving quality-preserving positron acceleration of small emittance beams, a critical requirement for the development of a linear electron-positron collider.

\section{Data availability}
\label{sec:data_availability}
All data in this study was obtained via the open-source, 3D, quasi-static PIC code HiPACE\texttt{++}, available on \href{https://github.com/Hi-PACE/hipace/}{this GitHub repository}. The input scripts for the HiPACE\texttt{++} simulations are openly available online~\cite{Diederichs:2023:dataset2}.

\section*{Acknowledgements}
We acknowledge support from DESY (Hamburg, Germany), a member of the Helmholtz Association HGF, and funding by the Helmholtz Matter and Technologies Accelerator Research and Development Program. This research was supported in part through the Maxwell computational resources operated at DESY, Hamburg, Germany. We gratefully acknowledge the Gauss Centre for Supercomputing e.V. (www.gauss-centre.eu) for funding this project by providing computing time through the John von Neumann Institute for Computing (NIC) on the GCS Supercomputer JUWELS at J\"ulich Supercomputing Centre (JSC).
This work was supported by the Director, Office of Science, Office of High Energy Physics,
of the U.S. Department of Energy under Contract No. DE-AC02-05CH11231,
and used the computational facilities at the National Energy Research Scientific Computing Center (NERSC).

\section*{References}
\bibliography{iopart-num-demo,references,references-ad}

\end{document}